\documentclass{elsart}
\begin{document}
\begin{frontmatter}
%
%preprint number
%
\[ \vspace{-6cm} \]
\noindent\hfill\hbox{\rm \normalsize{Alberta Thy 02-03}} \vskip 20pt
\title{Optimal Jet Finder}
\author{D. Yu. Grigoriev}
\address{Mathematical Physics, Natl. Univ. of Ireland Maynooth,
Maynooth, Co. Kildare, Ireland and Institute for Nuclear Research of RAS,
Moscow 117312, Russia}
\author{E. Jankowski}
\address{Department of Physics, University of Alberta,
Edmonton, AB, T6G 2J1, Canada}
\author{F. V. Tkachov}\footnote{Corresponding author: 
ftkachov@ms2.inr.ac.ru}
\address{Institute for Nuclear Research of RAS,
Moscow 117312, Russia}
\begin{abstract}
We describe a FORTRAN 77 implementation of the optimal jet definition
for identification of jets in hadronic final
states of particle collisions. 
We discuss details of the implementation, explain
interface subroutines and provide a usage example.
The source code is available from http://www.inr.ac.ru/$\sim$
ftkachov/projects/jets/
\end{abstract}
\end{frontmatter}
\footnotesize
\emph{Keywords:} hadronic jets, jet finding algorithms\\ 
\emph{PACS:} 13.87.-a, 29.85.+c
\normalsize
\newpage
\textbf{PROGRAM SUMMARY}\\ \\
\footnotesize
\emph{Title of program:} Optimal Jet Finder (OJF\_014)\\ \\
\emph{Catalogue identifier:} (supplied by the Publisher)\\ \\
\emph{Distribution format:} (supplied by the Program Library)\\ \\
\emph{Computer:}
any computer with the FORTRAN 77 compiler\\ \\
\emph{Tested with:}
g77/Linux on Intel, Alpha and Sparc;
Sun f77/Solaris (thwgs.cern.ch);
xlf/AIX (rsplus.cern.ch);
MS Fortran PowerStation 4.0/Win98\\ \\
Programming language used: FORTRAN 77\\ \\
\emph{Memory required:} $\sim$1 MB (or more, depending on the settings)\\ \\
\emph{Number of bytes in distributed program, including examples
and test data:} 1.39 MB\\ \\
\emph{Keywords:} hadronic jets; jet finding algorithms\\ \\
\emph{Nature of physical problem}\\
Analysis of hadronic final states in high energy particle collision experiments
often involves identification of hadronic jets. A large number of hadrons
observed in the detector is reduced to a few jets by means of a
\emph{jet finding algorithm}. The jets are used in further analysis which
would be difficult or impossible when applied directly to the hadrons.
Reference [1] provides a brief introduction to the subject of jet finding
algorithms and a general review of the physics of jets can be found
in [2].\\ \\
\emph{Method of solution}\\ 
The software we provide is an implementation of the so-called
\emph{optimal jet definition (OJD)}. The theory of OJD was developed in
[3], [4], [5]. The desired jet configuration is obtained as the one
that minimizes $\Omega_R$, a certain function of the input particles and
jet configuration.\\ \\
\emph{Restrictions on the complexity of the program}\\
The size of the largest data structure the program uses is
(maximal number of particles in the input) $\times$
(maximal number of jets in the output) $\times$ 8 bytes.
(For the standard settings $<$1 MB).
Therefore, there is no memory restriction for any conceivable application 
for which the program was designed.\\ \\
\emph{Typical running time}\\
The running time depends strongly on the physical process being analyzed
and the parameters used. For the benchmark process we studied,
$\mathrm{e}^+\mathrm{e}^-\rightarrow\mathrm{W}^+\mathrm{W}^-
\rightarrow\mathrm{4\;jets}$, with the average number of $\sim$ 80
particles in the input, the running time was $<\!10^{-2}$~s
on a modest PC per event with $n_\mathrm{tries}$=1.
(We took $n_\mathrm{tries}$=10 for all events in the benchmark process,
however for the majority of events $n_\mathrm{tries}\sim3$ would suffice.
For small values of $n_\mathrm{tries}$ the global minimum of $\Omega_R$ 
may be missed in some fraction of events resulting in the deterioration
of the precision of measurements based on the jet algorithm.)
For a fixed number of jets the complexity of the algorithm grows linearly
with the number of particles (cells) in the input, in contrast with other
known jet finding algorithms for which this dependence is cubic.
The reader is referred to [1] for a more detailed discussion of this issue.
\\ \\
\emph{References}\\
$[1]$ D.\ Yu.\ Grigoriev, E.\ Jankowski, F.\ V.\ Tkachov,
e-print hep-ph/0301185, to be published.\\
$[2]$ R.\ Barlow, Rep.\ Prog.\ Phys.\ \textbf{36}, 1067 (1993).\\
$[3]$ F.\ V.\ Tkachov, Phys.\ Rev.\ Lett.\ \textbf{73}, 2405 (1994);
Erratum, \textbf{74}, 2618 (1995).\\
$[4]$ F.\ V.\ Tkachov, Int.\ J.\ Mod.\ Phys.\ \textbf{A12}, 5411 (1997).\\
$[5]$ F.\ V.\ Tkachov, Int.\ J.\ Mod.\ Phys.\ \textbf{A17}, 2783 (2002).
\normalsize
\newpage
\tableofcontents
\newpage
\section{Introduction}
This paper represents an official public release of 
the Optimal Jet Finder (OJF) library --- 
a FORTRAN 77 implementation of the so-called \emph{optimal jet definition} (OJD)
for identification of jets in hadronic final
states of particle collisions.
For a brief introduction into the subject the reader is referred to
\cite{GJT},
and a general review of the physics of jets can be found in
\cite{barlow}.
The theory of OJD was developed in \cite{PRL}, \cite{whatisajet}, \cite{def},
to which papers the reader is referred for a detailed discussion
of the physical motivations behind and the derivation of OJD.
The principal points of the theory are summarized below.

(i) Calorimetric measurements with hadronic final states $\mathbf{P}$
must rely on observables $f\left(\mathbf{P}\right)$
that possess a special "calorimetric", or $C$-continuity which is
a non-perturbative generalization of the familiar IR safety
(see \cite{whatisajet} for details) and which guarantees a stability
of $f\left(\mathbf{P}\right)$ against distortions of $\mathbf{P}$
such as caused by detectors. Ref.\ \cite{whatisajet} pointed out
$C$-continuous analogues for a variety of observables usually studied via
intermediacy of jet algorithms.
The fundamental role of such observables is highlighted by two facts:
(1) An observable inspired by \cite{whatisajet} played an important role in
the selection of top quark events in the fully hadronic channel at D0
\cite{r19}, \cite{r20}.
(2) The Jet Energy Flow project \cite{r21} provides numerical evidence that
$C$-continuous observables may indeed help to go beyond the intrinsic
limitations of conventional procedure based on jet algorithms in the quest
for the 1\% precision level in the physics of jets.

(ii) The proposition that the observed event $\mathbf{P}$ inherits
information (as measured by calorimetric detectors) from the underlying
quark-and-gluon event $\mathbf{q}$ is expressed as 
%
%%
%%%
\begin{equation}\label{intro-1}
f\left(\mathbf{q}\right)
\approx
f\left(\mathbf{P}\right)
\mathrm{\quad for \;any \;}C\mathrm{\!-\!continuous}\;f.
\end{equation}
%%%
%%
%

(iii) For each parameter $M$ on which the probability distribution
$\pi_M\left(\mathbf{P}\right)$ of the observed events $\mathbf{P}$ may depend,
there exists an optimal observable
$f_\mathrm{opt} \left(\mathbf{P}\right) =
\partial_M \ln \pi_M \left(\mathbf{P}\right)$
for the best possible measurement of $M$ \cite{r18}. 
This is a reinterpretation of the Rao-Cramer
inequality and the maximal likelihood method of mathematical statistics
in terms of the method of moments. In the context of multi-hadron final
states as "seen" by calorimetric detectors,
such an observable is automatically $C$-continuous.

(iv) If the dynamics of hadronization is such that eq.\ (\ref{intro-1}) holds,
then good approximations for $f_\mathrm{opt}$ could exist among functions
that depend only on $\mathbf{Q}$ which is a parameterization
of $\mathbf{P}$ in terms of a few pseudo-particles
(jets), found from a condition modeled after eq.\ (\ref{intro-1}):
%
%%
%%%
\begin{equation}\label{intro-2}
f\left(\mathbf{Q}\right) \approx f\left(\mathbf{P}\right)
\mathrm{\quad for \;any \;}C\mathrm{\!-\!continuous}\;f.
\end{equation}
%%%
%%
%
This simply translates the meaning of jet finding as an inversion of
hadronization into the language of $C$-continuous observables.

(v) $C$-continuous observables can be approximated by sums of products
of simplest such observables that are linear in particles'
energies $E_a$: 
%
%%
%%%
\begin{equation}\label{intro-3}
\textstyle f\left(\mathbf{P}\right) =
\sum_a E_a f\left(\mathbf{\hat{p}}_a\right),
\end{equation}
%%%
%%
%
where $a$ runs over all particles in the event and $\mathbf{\hat{p}}_a$
(a unit vector) denotes the direction of the 3-momentum of the $a$-th
particle; $f$ is any continuous function of a direction only.
(The relevant theorems can be found in refs.\ \cite{whatisajet}
and \cite{def}.)

(vi) So it is sufficient to explore the criterion (\ref{intro-2}) with only
$f$'s of the form (\ref{intro-3}). Then one can perform a Taylor expansion
in angular variables and obtain a factorized bound of the form 
%
%%
%%%
\begin{equation}\label{intro-4}
\left| f\left(\mathbf{P}\right) - f\left(\mathbf{Q}\right) \right|
\le C_{f,R}
\times \Omega_R \left[\mathbf{P},\mathbf{Q}\right],
\end{equation}
%%%
%%
%
where all the dependence on $f$ is localized within $C_{f,R}$ 
(so the bound remains valid for any $C$-continuous $f$) and 
$\Omega_R \left[\mathbf{P},\mathbf{Q}\right]$ is a function
of the jet configuration $\mathbf{Q}$ (and the event $\mathbf{P}$)
only (the explicit form is given in section \ref{sec-def}).

(vii) Since the collection of values of all $f$ on a given event
$\mathbf{P}$ is naturally interpreted as the event's physical information
content, the bound (\ref{intro-4}) means that the distortion of such content
in the transition from $\mathbf{P}$ to $\mathbf{Q}$ can be controlled
by a single function; so the loss of physical information in the transition
is minimized if $\mathbf{Q}$ corresponds to the global minimum of $\Omega_R$.
The Optimal Jet Definition amounts to finding the jet configuration
$\mathbf{Q}$ which minimizes $\Omega_R \left[\mathbf{P},\mathbf{Q}\right]$,
depending on specific application, either with a given number
of jets or with a minimum number of jets while satisfying the restriction 
$\Omega_R \left[\mathbf{P},\mathbf{Q}\right]<\omega_\mathrm{cut}$ 
with some parameter $\omega_\mathrm{cut}>0$ which is similar to the jet
resolution $y_\mathrm{cut}$ of recombination algorithms.

The beta versions of the OJF library have been available since 1999
(see \cite{GrigorievTkachov}).
The publicly available code \cite{r25} was first developed in the 
programming language Component Pascal \cite{r26} from 
the highly regarded Pascal/Modula-2/Oberon family, featuring a unique
combination of intellectual manageability, safety, and efficiency
of compiled code (for more on this see \cite{fvtcpc2001}).
This made possible the experimentation needed
to find a working and correct algorithm. 
Only after that the final port to FORTRAN 77 was performed. 
A subsequent testing on a total of $O(10^{10})$ events \cite{r27} 
and a substantially
independent verification \cite{r28} revealed no defects 
of significance, indicating a high reliability of the code as a result
of the adopted development process. 

The OJF library can be used to obtain OJD implementations adapted for
specific applications (see below).

A typical program based on OJF takes as an input the collection of particles 
(or detector cells), characterizing them by energies and directions.
Then the optimal jet configuration is found by minimizing some function
of the particles and jet configuration.
Each particle can be shared among jets (in conventional schemes
only an entire particle can belong to a single jet), so
for an input event containing $n$ particles and resulting in $N$
jets, the number of degrees of freedom is $n\cdot N$.
A typical application (for instance, analysis of data for LHC) may
require $n\sim 100...400$ and $N\sim 6$ and therefore
$n\cdot N\sim O\left(1000\right)$.
A large number of degrees of freedom makes the minimization problem
non-trivial. The program we describe implements a practical
algorithm 
(taking a fraction of a second per event on a modest personal computer)
solving the optimization problem inherent in OJD. 
A key subroutine performing the minimization is
\texttt{Q\_minimize}.

For a fixed number of jets the computational complexity of the algorithm
grows as the first power of the number of particles in the input,
$O\left(n\right)$ (see \cite{GJT} for the corresponding empirical data;
for comparison, the computational complexity is $O\left(n^3\right)$ 
for the kT algorithm \cite{kT}, \cite{KTCLUS}.)
This feature makes OJF especially attractive for processing
events with many calorimetric cells in the input.
For a benchmark process we studied,
$\mathrm{e}^+\mathrm{e}^-\rightarrow\mathrm{W}^+\mathrm{W}^-
\rightarrow\mathrm{4\;jets}$,
OJF started to become more time efficient than the standard kT
roughly at 90 cells in the input.\footnote{For $n_\mathrm{ntries}=1$,
see section \ref{sec-def}.}

The version of OJF being described is called OJF\_014.
The FORTRAN 77 source code is available from http://www.inr.ac.ru/$\sim$
ftkachov/projects/jets/

OJF contains many interface subroutines
(usually with names starting with \texttt{set\_} or \texttt{get\_})
allowing to input data, read output and change the algorithm parameters.
The user is not supposed to write data directly to common blocks
but is to use the interface subroutines. Similarly, only the parameters
that can be accessed by the user callable subroutines are supposed
to be changed (at least at level of standard application).

OJF has been compiled on various platforms:
g77/Linux on Intel, Alpha and Sparc,
Sun f77/Solaris (thwgs.cern.ch),
xlf/AIX (rsplus.cern.ch),
MS Fortran PowerStation 4.0/Win98.
Its numerical stability was tested on a total of $\sim 10^{10}$ events.

OJF contains options to handle both the center of mass (spherical) kinematics
of lepton collisions 
and the cylindrical kinematics of hadron collisions.
%--------------------------------------------
\section{Algorithm}
%
%
%%%%%%%%%%%%%%%%%%%%%%%%%%%%%%%%%%%%%%%%%%%%%
%
%
\subsection{Criterion for the optimal jet configuration}
\label{sec-def}
%
%
% input
%
%
The input for the program is \emph{an event}: a collection of $n$
\emph{particles} from the detector (or $n$ hit detector cells),
indexed with $a=1,2,3,...,n.$
Each particle is characterized by its energy, $E_a$, and its direction
described by the standard angles $\theta _{a}$, $\varphi _{a}$
or equivalently by transverse energy, $E^{\perp }_{a}$,
pseudorapidity, $\eta_{a}$, and the angle $\varphi _{a}$.
The $a$-th particle in the input is assigned the 4-momentum $p_{a}$:
%
%p_a
%
\begin{equation}
\label{pa-1}
p_{a}=E_{a}\cdot\left(1,\,
\sin \theta _{a}\cos \varphi_{a},\,
\sin \theta _{a}\sin \varphi_{a},\,
\cos \theta _{a}\right) 
\end{equation}
or
%
%p_a
%
\begin{equation}
\label{pa-2}
p_{a}=E^{\perp }_{a}\cdot\left(
\cosh \eta _{a},\,
\cos\varphi _{a},\,
\sin \varphi _{a},\,
\sinh \eta _{a}\right).
\end{equation}
depending on what parameters describe the particles.
%
%
% output
%
%
The output of the program is a set of $N$ jets,
indexed with $j=1,2,3,...,N$. The jet configuration is described
by \emph{recombination matrix} $\left\{z_{aj}\right\}$ components
of which satisfy: 
%
%recombination matrix
%
\begin{equation}
\label{z-1}
0\leq z_{aj}\leq 1\quad\mathrm{for\quad all}\quad a,j,
\end{equation}
%
%recombination matrix
%
\begin{equation}
\label{z-2}
\sum ^{N}_{j=1}z_{aj}\leq 1\quad\mathrm{for\quad all}\quad a.
\end{equation}
The number $z_{aj}$ gives the fraction of the $a$-th particle which goes
into formation of the $j$-th jet. Each $z_{aj}$ can take any value
between 0 and 1. The final value of the recombination matrix
$\left\{z_{aj}\right\}$ is the result of the algorithm.
The 4-momentum $q_{j}$ of the $j$-th jet is defined as
\begin{equation}
\label{qj-1}
q_{j}=\sum ^{n}_{a=1}z_{aj}p_{a}.
\end{equation}

The final (optimal) jet configuration is the one that
minimizes the value of some function
$\Omega\left(\left\{z_{aj}\right\}\right)$
depending on the recombination matrix $\left\{z_{aj}\right\}$
and all $p_a$ as parameters.

The definition of $\Omega$ follows some intermediate sub-definitions.
Part of the $a$-th particle that do not go into formation of any jet: 
\begin{equation}
\label{z-bar}
\overline{z}_{a}\equiv 1-\sum ^{N}_{j=1}z_{aj}.
\end{equation}
The rest of the definitions are given separately for
spherical kinematics (lepton collisions) and for cylindrical
kinematics (hadron collisions).
\paragraph*{Spherical kinematics.}
Overall energy left outside jets $E_{\mathrm{soft}}$,
called \textit{soft energy}:
%
%soft energy cm kinematics
%
\begin{equation}
\label{soft-e-cm}
E_{\mathrm{soft}}\equiv \sum ^{n}_{a=1}\overline{z}_{a}E_{a}.
\end{equation}
The function $Y$, called \textit{fuzziness}:
%
%fuzziness
%
\begin{equation}
\label{fuziness}
Y\equiv 2\sum ^{N}_{j=1}q_{j}\widetilde{q_{j}},
\end{equation}
where $\widetilde{q_{j}}$ is light-like ($\widetilde{q_{j}}^{2}=0$)
4-direction defined:
\begin{equation}
\label{q-tylda-cm}
\widetilde{q_{j}}\equiv
\left(1,\,
\sin\theta_j\cos\varphi _{j},\,
\sin\theta_j\sin\varphi _{j},\,
\cos\theta_j\right),
\end{equation}
with
%
%cos-theta
%
\begin{equation}
\label{cos-theta}
\cos\theta_j\equiv\frac
{\left(q_{j}\right)_{z}}
{\sqrt{\left( q_{j}\right)^{2}_{x}
       +\left(q_{j}\right)^{2}_{y}
       +\left(q_{j}\right)^{2}_{z}}},
\end{equation}
%
%cos-phi
%
\begin{equation}
\label{cos-phi}
\cos\varphi_{j}\equiv\frac
{\left(q_{j}\right)_{x}}
{\sqrt{\left( q_{j}\right)^{2}_{x}
      +\left( q_{j}\right)^{2}_{y}}},
\end{equation}
%
%sin-phi
%
\begin{equation}
\label{sin-phi}
\sin \varphi _{j}\equiv\frac
{\left( q_{j}\right) _{y}}
{\sqrt{\left( q_{j}\right)^{2}_{x}
      +\left( q_{j}\right) ^{2}_{y}}}.
\end{equation}
\paragraph*{Cylindrical kinematics.}
The soft energy is the overall \emph{transverse} energy left
outside the jets
\begin{equation}
\label{e-soft-cyl}
E_{soft}\equiv\sum^n_{a=1}\overline{z}_{a}E_{a}^{\perp}.
\end{equation}
The fuzziness $Y$ is defined again by
(\ref{fuziness}) with $\widetilde{q_{j}}$, light-like
($\widetilde{q_{j}}^{2}=0$) 4-direction given by:
\begin{equation}
\label{q-tylda-cyl}
\widetilde{q_{j}}\equiv
\left(
\cosh\eta_j,
\cos\varphi _{j},
\sin\varphi _{j},
\sinh\eta_j
\right),
\end{equation}
where
%
%eta
%
\begin{equation}
\label{eta}
\eta_j\equiv\frac
{\sum ^{n}_{a=1}z_{aj}E_{a}^{\perp}\eta _{a}}
{\sum ^{n}_{a=1}z_{aj}E_{a}^{\perp}},
\end{equation}
and $\cos\varphi_j$, $\sin\varphi_j$
given by (\ref{cos-phi}), (\ref{sin-phi}).

Finally, \textbf{in both cases}, $\Omega$ is a linear combination of $Y$
and $E_{\mathrm{soft}}$ with the parameter $R$ weighting their relative
contribution:
\begin{equation}
\label{def-omega}
\Omega\left(\left\{z_{aj}\right\}\right)\equiv
\frac{1}{R^{2}}Y+E_{soft}.
\end{equation}

If we intend to reconstruct an event to some fixed number of jets
the procedure is as follows. Start with some initial value of
the recombination matrix $z_{aj}$, for example chosen randomly
and minimize the function $\Omega\left(\left\{z_{aj}\right\}\right)$
with respect to $\{z_{aj}\}$. The algorithm described in section
\ref{sec-algo}
allows one to find a local minimum while starting with some initial
value of $\{z_{aj}\}$. Repeat the procedure a few times
(parameter $n_{\mathrm{tries}}$), starting each time
with different initial value of $\{z_{aj}\}$
and take the smallest of the minima obtained at each try.
The value of the recombination matrix $\{z_{aj}\}$ that gives
the smallest of the minima is the final jet configuration.
If the initial value of $\{z_{aj}\}$ is not chosen
randomly it is useless to do the minimization procedure more than once
as the minimization algorithm is deterministic.

If the number of jets is to be determined in the process
of jet reconstruction we can repeat the procedure described above
for different number of jets $N$ each time. This means, we start
with some minimal number of jets, e.g. $N=1$ and find the corresponding
$N$-jet configuration and check whether
\begin{equation}
\label{criterion}
\Omega\left(\left\{z_{aj}\right\}\right)<\omega_{\mathrm{cut}}
\end{equation}
is fulfilled for that configuration. If so, this is the \emph{final}
jet configuration we look for. If not, we increase $N$ by one
and repeat the $N$-jet procedure for the new $N$. We continue until
(\ref{criterion}) is finally satisfied (which has to be true for
some sufficiently large number of jets).
The parameter $\omega_{\mathrm{cut}}$ is some (small) positive number
one chooses and it is analogous to the jet resolution parameter
of conventional recombination algorithms.
%
%
%
%%%%%%%%%%%%%%%%%%%%%%%%%%%%%%%%%%%%%%%%%%%%%%%%%%%%%%%%
%
%
%
\subsection{Algorithm for minimizing $\Omega$}
\label{sec-algo}
The domain of the function $\Omega\left(\left\{z_{aj}\right\}\right)$
is a $\left(n\cdot N\right)$-dimensional product of simplices.
That is, for fixed $a$ the numbers $z_{aj}$, $j=1,...,N$,
satisfying conditions (\ref{z-1}), (\ref{z-2}) define an
$N$-dimensional simplex. In typical application
$n\sim 200$ (or more) and $N\sim 5$ and therefore $\Omega$
is a function of $\sim 1000$ variables.
The algorithm described below allows for efficient minimization of
$\Omega\left(\left\{z_{aj}\right\}\right)$.

The algorithm iteratively descends into local minimum of
$\Omega\left(\left\{z_{aj}\right\}\right)$
starting from a given initial value of $\left\{z_{aj}\right\}$.
At each iteration, subsequently for each particle,
$\left\{z_{aj}\right\}$
is moved into new position in the domain that gives the smaller
value of $\Omega$. The iteration loop is terminated when
no particle is moved at a single iteration, meaning that the
local minimum has been found. (Or some safe number of maximal
iterations has been exceeded.)

We describe now in detail how $\left\{z_{aj}\right\}$ is moved
in a single iteration step for a given particle.
Denote
$\mathbf{z}\equiv\left(z_1,z_2,...,z_N\right)$ with $z_j=z_{aj}$
and
$\Omega\left(\mathbf{z}\right)\equiv\Omega\left(\left\{z_{aj}\right\}\right)$
with fixed $a$ in both definitions.
The change in $\Omega$ when we change $\mathbf{z}$ to
$\mathbf{z}+\tau \mathbf{d}$ can be described locally as
%
%change-in-omega
%
\begin{equation}
\label{new-om}
\Omega\left(\mathbf{z}+\tau\mathbf{d}\right)=
\Omega\left(\mathbf{z}\right)+\tau\mathbf{f}\cdot\mathbf{d}
+O\left(\tau^2\right),
\end{equation}
where
$\mathbf{f}=\left(f_1,...,f_N\right),$
$f_j\equiv
\partial\Omega\left(\mathbf{z}\right)/
\partial z_j,$
$\mathbf{f}\cdot\mathbf{d}=\sum _{j=1}^N f_j d_j$
and
$\mathbf{d}=\left(d_1,...,d_N\right)$ describes some direction.
If $\mathbf{z}$ were not constrained to the simplex
we could take $\mathbf{d}=-\mathbf{f}$ and some $\tau>0$
to decrease $\Omega$. But choosing $\tau$ and $\mathbf{d}$
we have to ensure that $\mathbf{z}+\tau\mathbf{d}$ is within
the simplex. Rewrite
%
%fd
%
\begin{equation}
\label{fd}
\mathbf{f}\cdot\mathbf{d}=
\sum_{j=1}^N f_j d_j=
\sum_{j=1}^N \bar{f}_j d_j+\bar{f}_0 d_0
\end{equation}
with the following definitions
%
% definition of f_bar 
%
\begin{equation}
\label{def-f-bar}
\bar{f}_j\equiv f_j-f_J,
\quad 
\bar{f}_0 \equiv -f_J,
\end{equation}
%
%
%
%
% definition of d_0
%
\begin{equation}
\label{def-d-0}
d_0 \equiv -\sum_{j=1}^N d_j,
\end{equation}
where $J$ is any of $1,...,N$ for which $z_J>0$
(there always must be such $J$).
Now $\mathbf{d}$ can be chosen as follows
%
% choice of vector d
%
\begin{equation}
\label{vec-d}
d_j \equiv \left\{ \begin{array}{ll}
\max\left(0,-\bar{f}_j\right)
& \mathrm{for\;all}\;j=0,...,N,\;
\mathrm{for\;which}\; z_j=0 \\
-\bar{f}_j
& \mathrm{for\;all}\;j=0,...,N,\;
j\ne J\;\mathrm{for\;which}\;z_j>0
\end{array} \right.
\end{equation}
and $d_J$ is chosen so that (\ref{def-d-0}) is satisfied.
With such choice of $\mathbf{d}$ and the proper parameter $\tau$
the new candidate minimum $\mathbf{z}+\tau\mathbf{d}$
will belong to the simplex and
$\Omega\left(\mathbf{z}+\tau\mathbf{d}\right)<\Omega\left(\mathbf{z}\right).$
In the above prescription the choice of $J$ is arbitrary. We found
it advantageous to choose $J$ ($z_J>0$) such that the norm
\begin{equation}
\label{d-norm}
\left|\left(\mathbf{d},d_0\right)\right|\equiv
\max\left\{\left|d_j\right|:\; j=0,1,...,N\right\}
\end{equation}
is maximal. The choice of step length $\tau$ is determined
by the experimental finding that the minimum tends to be located
at the boundary of the simplex. We find
\begin{equation}
\label{step}
\tau=\min\left(\left\{
-\frac{z_j}{d_j}:\quad
j=0,...,N,\:z_j>0\;\mathrm{and}\;d_j<0
\right\}\right)
\end{equation}
from the requirement
that the new candidate minimum $\mathbf{z}+\tau\mathbf{d}$ should be
located at the boundary of the simplex and if this results in an
increase of the value of $\Omega$ then $\tau$ is iteratively
divided by a constant factor $\left(\sim 3\right)$ until minimum is found.

An important technical implementation detail is the so-called
``snapping''. If some $z_{aj}$ is small enough (i.e. $\mathbf{z}$
is close enough to a boundary of the simplex) then it is set to zero.
A similar snapping is used for the direction $\mathbf{d}$.
%
%
%
%
%%%%%%%%%%%%%%%%%%%%%%%%%%%%%%%%%%%%%%%%%%%%%%%%%%%%%%%%
%
%
%
\subsection{Formulas}
We give here the explicit formulas for derivatives
$f_j\equiv
\partial\Omega\left(\left\{z_{aj}\right\}\right)/
{\partial z_{aj}}$
used within the algorithm (derived from the definitions given
in the previous sections).
\paragraph*{Spherical kinematics:}
\begin{equation}
\label{f-j-spher}
f_j=2p_a\tilde{q}_j-E_a.
\end{equation}
\paragraph*{Cylindrical kinematics:}
\begin{equation}
\label{f-j-cyl}
f_j=2p_a\tilde{q}_j-
\frac{E^\perp_a}{\sum_{j=1}^N z_{aj} E^\perp_a}
\left(\eta_a-\eta_j\right)
\left(q_j^0\sinh\eta_j-q_j^\mathrm{z}\cosh\eta_j\right)
-E^\perp_a.
\end{equation}
%
%
%
%
%
%--------------------------------------------
%
\section{Code and usage}
We describe a FORTRAN 77 implementation of the algorithm explained
in the previous section.
\subsection{Code and data structure}
The code (file \texttt{ojf\_014.f}) consists of subroutines (and functions)
which can be divided
in three logical groups: (i) interface subroutines, (ii) core subroutines
and (iii) example jet search or result printing subroutines.
In addition block data \texttt{ojf\_lock} contains default values of some
program parameters.
The interface subroutines allow the user to enter input data,
to read output or already entered input,
to set or change program parameters and to obtain information
about current program parameters.
\emph{All parameters that are supposed to be set or changed by the user
can be accessed by these subroutines.
The same applies to all input and output data.}
The user is not supposed to write directly to common blocks.
The core subroutines (functions) perform
$\Omega\left(\left\{z_{aj}\right\}\right)$
minimization and
conversion between various data forms. The user is not supposed to call
them directly except for \texttt{Q\_minimize}. The subroutine
\texttt{Q\_search} is an example application of OJF frame
to simple jet search (see section \ref{sec-q-search}).
The user may want to modify it or write their own subroutines if needed.

All floating point variables within the program are defined as\\
\texttt{DOUBLE PRECISION}. If the user employs \texttt{REAL}
type variables they should ensure that a proper conversion 
of the parameter values is made in the calls
of the OJF subroutines.

The file \texttt{ojf\_com.fh} contains common block definitions
of internal data structures for OJF, for instance, matrices for parameters of
input particles, output jets parameters and recombination matrix
$\left\{z_{aj}\right\}.$
The file \texttt{ojf\_par.fh} contains the definitions of constants
used within the program.
The file \texttt{ojf\_kin.fh} contains the definitions of two constants:
\texttt{sphere=1, cylinder=2}. The file can be contained in user programs
whenever reference to kinematics type is made, e.g.\\
\begin{quote}
\texttt{...}\\
\texttt{INCLUDE 'ojf\_kin.fh'}\\
\texttt{INTEGER kinematics}\\
\texttt{...}\\
\texttt{kinematics=sphere}\\
\texttt{event\_setup\_begin( kinematics )}\\
\texttt{...}\\
\end{quote}
The other two files (\texttt{ojf\_com.fh} and \texttt{ojf\_kin.fh})
normally do not need to be contained in user programs.
\subsection{Normalization of energy units}
\label{normalization}
The energies $E_a$ or $E^{\perp }_{a}$ of input particles 
and the corresponding 4-momenta $p_a$ are normalized 
(after being entered) according to
%
% norm-1
%
\begin{equation}
\label{norm-1}
E_a\rightarrow\frac{E_a}{\sum_{a=1}^n E_a},
\quad
p_a\rightarrow\frac{p_a}{\sum_{a=1}^n E_a}
\end{equation}
for spherical kinematics or according to
%
% norm-2
%
\begin{equation}
\label{norm-2}
E^\perp_a\rightarrow\frac{E^\perp_a}{\sum_{a=1}^n E^\perp_a},
\quad
p_a\rightarrow\frac{p_a}{\sum_{a=1}^n E^\perp_a}
\end{equation}
for cylindrical kinematics.
The normalization constant $\sum_{a=1}^n E_a$ or $\sum_{a=1}^n E^\perp_a$
is stored to interpret properly the final output.
The normalization allows to make the implementation independent of energy
units and scale. 
\subsection{Error messages}
Significant part of the code consists of various checks.
For example:
\begin{quote}
\verb|...|\\
\verb|IF (.NOT. ojf_event_begin) THEN|\\
\verb|   WRITE(6,*) 'add_particle: 20: wrong call sequence'|\\
\verb|   WRITE(6,*) 'call event_setup_begin first'|\\
\verb|   STOP 'add_particle: 20'|\\
\verb|END IF|\\
\verb|...|
\end{quote}
The checks are used to assure that subroutines are not called
in inappropriate order, chosen parameters or input data do
not have pathological values and that the program runs properly.
The check can generate an error message and terminate the
program.
Messages with numbers 20-29 are due to the user errors.
Messages with numbers $\ge30$ are generated by program failures,
so should you get such a message, please inform the authors; 
please include the corresponding event in text form.
\subsection{Key minimization subroutine \texttt{Q\_minimize}}
\label{sec-q-min}
The subroutine \texttt{Q\_minimize} minimizes
$\Omega\left(\left\{z_{aj}\right\}\right)$
for a given number of jets 
starting from the existing configuration of
$\left\{z_{aj}\right\}.$ An example program that uses \texttt{Q\_minimize}
is given in section \ref{sec-example}.

The subroutine performs iteratively the minimization algorithm
described in section \ref{sec-algo}. The iteration loop is terminated
when no particle is moved in a single iteration or the maximal number
of iterations is exceeded. (We regard that the minimum is found only in
the former case.) Default value of the maximal number of iterations
is set 1000 which corresponds to $\sim$ 1 second of computing time on a
modest computer. It can be changed with \texttt{set\_maxiter ( maxiter )},
see section \ref{sec-set-param}. In each iteration, a loop over all
particles is run ($a=1,2,...,n$). For each particle separately, new
candidate $\left\{z_{aj}\right\}$ for the minimum is found. The direction,
$\mathbf{d},$ and step $,\tau,$ are computed according to the procedure
described in section \ref{sec-algo}.
Unless the step is zero or ``infinity'', indicating
that the particle should not be moved, the condition
\begin{equation}
\label{new-omega}
\Omega\left(\mathbf{z}+\tau\mathbf{d}\right)<
\Omega\left(\mathbf{z}\right).
\end{equation}
is checked.
If the condition is met the recombination matrix
$\left\{z_{aj}\right\}$ is moved into the new position.
If not, the step is reduced 3 times and (\ref{new-omega})
is checked again. If it is not true $\tau$ is reduced again and so on.
If $\tau$ falls below some small parameter
($\tau\left|\left(\mathbf{d},d_0\right)\right|\le\mathtt{eps\_dist}$)
the particle is not moved
and the program proceeds to the next particle.
\subsection{User callable subroutines}
We describe all user callable subroutines other than \texttt{Q\_minimize}
explained above.
\subsubsection{Event setup}
\label{sec-event-setup}
\paragraph*{\texttt{event\_setup\_begin ( kinematics )}}~\\
\begin{tabular}{lll}
\hline
\emph{input:} & &\\
\texttt{INTEGER} &  \texttt{kinematics} &
kinematics type\\
\hline
\end{tabular}\\
The subroutine begins initialization of a new event.
It must be called before event data is entered.
The parameter \texttt{kinematics}
informs the program what type of kinematics is used: spherical (center of mass
collisions), \texttt{kinematics}=1 or cylindrical (hadron collisions),
\texttt{kinematics}=2. If the file \texttt{ojf\_kin.fh} is included
constants \texttt{sphere} and \texttt{cylinder} can be used to assign
value to \texttt{kinematics}:
\begin{quote}
\texttt{...}\\
\texttt{INCLUDE 'ojf\_kin.fh'}\\
\texttt{INTEGER kinematics}\\
\texttt{...}\\
\texttt{kinematics=sphere}\\
\texttt{event\_setup\_begin( kinematics )}\\
\texttt{...}\\
\end{quote}
Kinematics ought to be set once for all events in a job.
\paragraph*{\texttt{add\_particle ( energy, theta, phi )}}~\\
\begin{tabular}{lll}
\hline
\emph{input:} & &\\
\texttt{DOUBLE PRECISION} &  \texttt{energy} &
energy $E_a$\\
\texttt{DOUBLE PRECISION} &  \texttt{theta} &
angle $\theta_a$\\
\texttt{DOUBLE PRECISION} &  \texttt{phi} &
angle $\phi_a$\\
\hline
\end{tabular}\\
The subroutine is used to enter input data.
It must be called between\\
\texttt{event\_setup\_begin}
and \texttt{event\_setup\_end}. 
Each call adds a particle (=detector cell) to the event.
The energy $E_a$ of the particle can be in any units,
for example GeV. The direction of the particle is 
described by the standard angles $\theta_a$ (measured from beam axis)
and $\phi_a$.
\paragraph*{\texttt{add\_particle\_raw ( px, py, pz )}}~\\
\begin{tabular}{lll}
\hline
\emph{input:} & &\\
\texttt{DOUBLE PRECISION} &  \texttt{px, py, pz} &
3-momentum components\\
\hline
\end{tabular}\\
The subroutine is used to enter input data,
as an alternative to \texttt{add\_particle}.
It must be called between \texttt{event\_setup\_begin}
and \texttt{event\_setup\_end}.
Each call adds a particle (=detector cell) to the event. 
The parameters \texttt{px}, \texttt{py}, \texttt{pz}
are 3-momentum components in the same units as energy in
\texttt{add\_particle}. The beam axis is in z-direction.
The subroutine is useful with output of Monte Carlo event generators.
It can be freely mixed with \texttt{add\_particle}.
\paragraph*{\texttt{event\_setup\_end}}~\\
must be called after all input particles are entered and
before the jet search can be undertaken.
No particles can be added to the event afterwards.
This subroutine is needed for internal housekeeping.
For instance, it provides the proper normalization of 
the energies of the particles.
\subsubsection{Setup of initial jet configuration}
\label{sec-jet-setup}
\paragraph*{\texttt{jets\_setup\_begin ( njets, Radius )}}~\\
\begin{tabular}{lll}
\hline
\emph{input:} & &\\
\texttt{INTEGER} &  \texttt{njets} &
number of jets, $N$\\
\texttt{DOUBLE PRECISION} &  \texttt{Radius} &
parameter $R$\\
\hline
\end{tabular}\\
The subroutine has to be called to begin setup of the initial jet
configuration - the initial value of the recombination matrix
$\left\{z_{aj}\right\}$, necessary for the
iterative minimization of $\Omega$, as explained in section
\ref{sec-algo}.
It is called automatically by \texttt{Q\_search} but
it must be called explicitly if \texttt{Q\_minimize} is used instead.
The number of jets, $N$, must be positive. (The event will be
reconstructed to the number of jets entered here.)
$R$ is the parameter
in equation (\ref{def-omega}). It has to be positive and not too close
to zero. The bigger R is, the less energy is left outside the jets.
New configurations of jets can be set up any number of times for the
same event.
The value of the seed from which the random number generator will start
for this jet configuration is stored at this point.
(From this point until the first invocation of anything random,
the seed can be reset by \texttt{set\_seed.})
If you need only to change $R$ and proceed with minimization 
starting from the current configuration, use \texttt{reset\_Radius}.
\paragraph*{\texttt{set\_seed ( seed )}}~\\
\begin{tabular}{lll}
\hline
\emph{input:} & &\\
\texttt{INTEGER} &  \texttt{seed} &
\\
\hline
\end{tabular}\\
This is to allow variation in random initial configurations of jets
in case there are several local minima.
It may be called once for a whole sequence of events - each event 
starts with a seed set up by the internal random number generator.
The seed can be read (see \texttt{get\_seed}) and used as a key to regenerate 
the corresponding configuration of jets (i.e. local minimum;
so the local minimum is completely determined by its seed).
It must be called after \texttt{jets\_setup\_begin} but cannot be called 
after the first invocation of \texttt{init\_z\_random} or
\texttt{init\_random\_all} or \texttt{jets\_setup\_end}
and until the next \texttt{jets\_setup\_begin}.
\paragraph*{\texttt{reset\_Radius ( Radius )}}~\\
\begin{tabular}{lll}
\hline
\emph{input:} & &\\
\texttt{DOUBLE PRECISION} &  \texttt{Radius} &
parameter $R$\\
\hline
\end{tabular}\\
The subroutine changes the value of
the parameter $R$ in equation (\ref{def-omega}). 
$R$ has to be positive and not too close to zero.
A large value of $R$ means less energy is left outside jets.
The subroutine can be called at any time
- the current configuration of jets is not affected
(only $\Omega$ is recalculated properly).
This may be useful for setting up interesting variations of the 
algorithm ("annealing") in which one starts from
some small value of $R$ and then changes it gradually, 
fine-tuning the resulting jet configurations by calls to
\texttt{Q\_minimize}.
With infinitesimal values of $R$, the global minimum 
occurs for jet configurations with the most energetic particles 
playing the role of jets, so this can be used to obtain 
the most energetic (narrow clusters of) particles.
\paragraph*{\texttt{init\_z\_random\_all}}~\\
The subroutine can only be called between \texttt{jets\_setup\_begin}
and\\
\texttt{jets\_setup\_end}.
It is the simplest way to initialize the recombination matrix
$\left\{z_{aj}\right\}$: 
completely and uniformly random in the direct product of all the
simplices corresponding to particles.
If only specific particles need to be randomized,
\texttt{init\_z\_random( a )} should be used.
If only specific particles need to be set non-randomly, 
\texttt{init\_z( a, z\_in)} or \texttt{assign\_to\_jet( a, j )}
should be called for those particles. 
Then \texttt{init\_z\_random\_all} can be called to randomize
the remaining particles.
If this is not called explicitly, the particles not explicitly initialized
are set to "neutral" positions (democratically shared between all jets 
and the soft energy).
\paragraph*{\texttt{assign\_to\_jet ( a, j )}}~\\
\begin{tabular}{lll}
\hline
\emph{input:} & &\\
\texttt{INTEGER} &  \texttt{a} &
index of the particle\\
\texttt{INTEGER} &  \texttt{j} &
index of the jet\\
\hline
\end{tabular}\\
The subroutine can only be called between \texttt{jets\_setup\_begin}
and\\ \texttt{jets\_setup\_end}.
It can be used to set the initial configuration of jets
explicitly, for instance, when the output of another jet algorithm
is to be fine-tuned.
It sets the value $z_{aj}=1$ for the given $a$, $j$, i.e.
directly assigns the $a$-th particle to the $j$-th jet.
It must have $1\le a \le n$ and $0\le j \le N$;
$j=0$ corresponds to soft energy.
The subroutine only sets the initial configuration.
No elements of the recombination matrix are protected
from being changed by subsequent minimizations.
\paragraph*{\texttt{init\_z\_from ( a, z\_in )}}~\\
\begin{tabular}{lll}
\hline
\emph{input:} & &\\
\texttt{INTEGER} &  \texttt{a} &
index of the particle\\
\texttt{DOUBLE PRECISION} &  \texttt{z\_in(0:njets\_max)} &
components $z_{a0}$, $z_{a1}$, ..., $z_{aN}$\\
\hline
\end{tabular}\\
The subroutine can only be called between \texttt{jets\_setup\_begin}
and\\ \texttt{jets\_setup\_end}.
It can be used to set the initial configuration of jets explicitly,
for instance, to fine-tune the output of another jet algorithm.
It initializes the recombination matrix $\left\{z_{aj}\right\}$
for the $a$-th particle, i.e. sets $z_{a0}$, $z_{a1}$, ..., $z_{aN}$.
Only $N+1$ components of the vector \texttt{z\_in(0:njets\_max)} are used.
The components must be all non-negative but do not need to be normalized 
correctly - correct normalization will be imposed automatically;
\texttt{z\_in(0)} is the particle's fraction relegated to soft energy.
For instance, \texttt{z\_in(j)} can be some measure of distance 
between the $a$-th particle and the $j$-th jet from another jet algorithm.
The subroutine only sets the initial configuration.
No elements of the recombination 
matrix are protected from being changed by subsequent minimizations.
\paragraph*{\texttt{init\_z\_random ( a )}}~\\
\begin{tabular}{lll}
\hline
\emph{input:} & &\\
\texttt{INTEGER} &  \texttt{a} &
index of the particle\\
\hline
\end{tabular}\\
The subroutine can only be called between \texttt{jets\_setup\_begin}
and\\ \texttt{jets\_setup\_end}.
It does random initialization of the recombination matrix
$\left\{z_{aj}\right\}$ for the $a$-th particle.
\paragraph*{\texttt{jets\_setup\_end}}~\\
The subroutine must be called prior to minimization. It does housekeeping
such as 
initialization of the particles whose recombination matrix elements 
have not been explicitly initialized by calls from
\texttt{init\_z\_random}.
\subsubsection{Setting algorithm control parameters}
\label{sec-set-param}
\paragraph*{\texttt{set\_maxiter ( maxiter )}}~\\
\begin{tabular}{lll}
\hline
\emph{input:} & &\\
\texttt{INTEGER} &  \texttt{maxiter} &
maximal number of iterations\\
\hline
\end{tabular}\\
The subroutine can be called to change the maximal number of iteration,
see \ref{sec-q-min}. Default value of the maximal number of iterations
is set to 1000 which corresponds to $\sim$ 1 second of computing time on
a modest PC. It can be called at any time.
\paragraph*{\texttt{set\_njets\_limits ( nstart, nstop )}}~\\
\begin{tabular}{lll}
\hline
\emph{input:} & &\\
\texttt{INTEGER} &  \texttt{nstart} &
starting number of jets\\
\texttt{INTEGER} &  \texttt{nstop} &
maximal number of jets\\
\hline
\end{tabular}\\
The subroutine is needed in conjunction with \texttt{Q\_search} only.
It sets the starting and the final number of jets in \texttt{Q\_search}
(see the end of section \ref{sec-algo}
and description of \texttt{Q\_search} in section \ref{sec-q-search}).
The parameters must obey $1\le\mathtt{nstart}\le\mathtt{nstop}$
and $\mathtt{nstop}\le\mathtt{njets\_max}$
(constant \texttt{njets\_max}, set in \texttt{ojf\_par.fh}, defines
the dimension of matrices and is the maximal allowed number of jets).  
The default values are: \texttt{nstart}=1 and \texttt{nstop}=
\texttt{njets\_max}=20.
The subroutine can be called any time.
\paragraph*{\texttt{set\_ntries ( n )}}~\\
\begin{tabular}{lll}
\hline
\emph{input:} & &\\
\texttt{INTEGER} &  \texttt{n} &
number of tries\\
\hline
\end{tabular}\\
The subroutine is needed in conjunction with \texttt{Q\_search} only.
It sets the number of tries to find the minimum with different random
initial configurations for each number of jets
(see the end of section \ref{sec-algo}
and description of \texttt{Q\_search}, section \ref{sec-q-search}).
The parameter \texttt{n} must be positive.
The larger \texttt{n}, the higher the probability that the found
configuration is the global minimum.
Note that number of local minima correlates positively 
with number of hard partons. Usually values $\sim10$ should suffice.
The subroutine can be called at any time.
\paragraph*{\texttt{set\_trace\_nmoved ( bool )}}~\\
\begin{tabular}{lll}
\hline
\emph{input:} & &\\
\texttt{LOGICAL} &  \texttt{bool} &
see text\\
\hline
\end{tabular}\\
The subroutine with parameter \texttt{bool=.TRUE.}
turns on the option in which \texttt{Q\_minimize} prints
how many particles were moved at each iteration;
with \texttt{bool=.FALSE.} it switches the option off (default).
The subroutine can be called any time.
\subsubsection{Access to parameters}
\paragraph*{\texttt{get\_kinematics ( kinematics )}}~\\
\begin{tabular}{lll}
\hline
\emph{output:} & &\\
\texttt{INTEGER} &  \texttt{kinematics} &
type of kinematics\\
\hline
\end{tabular}\\
The subroutine returns the type of kinematics.
The possible values are 1 (spherical kinematics, center of mass
collisions) and 2 (cylindrical kinematics, hadron collisions),
which is equivalent to constants \texttt{sphere} and \texttt{cylinder}
if the header file \texttt{ojf\_kin.fh} is included (see also section
\ref{sec-event-setup}).
The subroutine cannot be called prior to the very first call of
\texttt{event\_setup\_begin}.
\paragraph*{\texttt{get\_nparts ( nparts, e\_scale )}}~\\
\begin{tabular}{lll}
\hline
\emph{output:} & &\\
\texttt{INTEGER} &  \texttt{nparts} &
number of particles in the event\\       
\texttt{DOUBLE PRECISION} & \texttt{e\_scale} &
total energy of the event\\
\hline
\end{tabular}\\
The subroutine returns the number of particles, $n$, and the total energy
in the event.
The total energy is the sum of the usual energies of the particles
for spherical kinematics $\sum_{a=1}^n E_a$
and the sum of transverse energies for cylindrical kinematics
$\sum_{a=1}^n E^\perp_a$ in in physical units, i.e.
prior to the normalization $\sum_{a=1}^n E_a=1$ or $\sum_{a=1}^n E^\perp_a=1$.
In other words, \texttt{e\_scale} is the normalization constant.
Energy/momentum parameters returned by some other subroutines are normalized
by the value of \texttt{e\_scale}.
The subroutine cannot be called between
\texttt{event\_setup\_begin} and \texttt{event\_setup\_end}.
\paragraph*{\texttt{get\_particle ( a, e, xta, phi, p, ephys, pphys )}}~\\
\begin{tabular}{lll}
\hline
\emph{input:} & &\\
\texttt{INTEGER} &  \texttt{a} &
index of the particle\\
\hline
\emph{output:} & &\\
\texttt{DOUBLE PRECISION} & \texttt{e} &
normalized energy $E_a$ or $E^\perp_a$\\       
\texttt{DOUBLE PRECISION} & \texttt{xta} &
angle $\theta_a$ or pseudorapidity $\eta_a$\\
\texttt{DOUBLE PRECISION} & \texttt{phi} &
angle $\phi_a$\\
\texttt{DOUBLE PRECISION} & \texttt{p(0:3)} &
normalized 4-momentum $p_a$\\
\texttt{DOUBLE PRECISION} & \texttt{ephys} &
energy $E_a$ or $E^\perp_a$ not normalized\\
\texttt{DOUBLE PRECISION} & \texttt{pphys(0:3)} &
4-momentum $p_a$ not normalized\\
\hline
\end{tabular}\\
The subroutine returns parameters of the $a$-th particle.
For spherical kinematics the parameters are the usual energy, $E_a$,
and the standard angles $\theta_a$ (from the beam axis) and $\phi_a$.
For cylindrical kinematics the parameters are the transverse
energy, $E^\perp_a$, pseudorapidity, $\eta_a$, and the angle $\phi_a$.
The value of \texttt{e} is normalized and the \texttt{ephys} is in the
same units as used in input,
i.e. $\mathtt{ephys}=\mathtt{e}\cdot\mathtt{e\_scale}$ (see the previous
subroutine for \texttt{e\_scale}). All angles are in degrees.
In both kinematics, \texttt{p(0:3)} and \texttt{pphys(0:3)} are
the normalized and non-normalized 4-momenta, $p_a$, of the particle.
The subroutine cannot be called between \texttt{event\_setup\_begin}
and \texttt{event\_setup\_end}.
\paragraph*{\texttt{get\_njets ( njets )}}~\\
\begin{tabular}{lll}
\hline
\emph{output:} & &\\
\texttt{INTEGER} &  \texttt{njets} &
number of jets\\
\hline
\end{tabular}\\
The subroutine returns the number of jets in the current configuration of jets.
It cannot be called before the first configuration of jets is setup.
\paragraph*{\texttt{get\_seed ( seed )}}~\\
\begin{tabular}{lll}
\hline
\emph{input:} & &\\
\texttt{INTEGER} &  \texttt{seed} &
seed for random generator\\
\hline
\end{tabular}\\
The subroutine returns the value of the seed for the random generator,
used for setting up the current random jet configuration.
The value of the seed is ``locked''
(causing attempts to reset it to result in program termination)
by the first invocation of anything ``random'' and retained 
until ``unlocked'' and reset by \texttt{jets\_setup\_begin}.
\paragraph*{\texttt{get\_Radius ( R )}}~\\
\begin{tabular}{lll}
\hline
\emph{output:} & &\\
\texttt{DOUBLE PRECISION} &  \texttt{R} &
parameter $R$ in eq. (\ref{def-omega})\\
\hline
\end{tabular}\\
The subroutine returns current value of the parameter $R$
in equation (\ref{def-omega}).
\paragraph*{\texttt{get\_maxiter ( maxiter )}}~\\
\begin{tabular}{lll}
\hline
\emph{output:} & &\\
\texttt{INTEGER} & \texttt{maxiter} &
maximal number of iteration\\
\hline
\end{tabular}\\
The subroutine returns the maximal number of iterations
(see section \ref{sec-q-min}).
\paragraph*{\texttt{get\_njets\_limits ( nstart, nstop )}}~\\
\begin{tabular}{lll}
\hline
\emph{output:} & &\\
\texttt{INTEGER} & \texttt{nstart} &
starting number of jets\\
\texttt{INTEGER} & \texttt{nstop} &
maximal number of jets\\
\hline
\end{tabular}\\
The subroutine returns the current values of the starting number of jets
and the maximal number of jets in the subroutine \texttt{Q\_search}
(see the end of section \ref{sec-algo}
and description of \texttt{Q\_search} in  section \ref{sec-q-search}).
\paragraph*{\texttt{get\_ntries ( n )}}~\\
\begin{tabular}{lll}
\hline
\emph{output:} & &\\
\texttt{INTEGER} &  \texttt{n} &
number of tries\\
\hline
\end{tabular}\\
The subroutine returns the current number of tries in \texttt{Q\_search},
the number of attempts to find minimum with different random initial
configurations for each number of jets (see the end of section \ref{sec-algo}
and description of \texttt{Q\_search}, section \ref{sec-q-search}).
\subsubsection{Access to results}
\label{sec-acc-res}
\paragraph*{\texttt{get\_criterion ( omega, y, esoft )}}~\\
\begin{tabular}{lll}
\hline
\emph{output:} & &\\
\texttt{DOUBLE PRECISION} &  \texttt{omega} &
value of $\Omega$\\
\texttt{DOUBLE PRECISION} &  \texttt{y} &
value of $Y$\\
\texttt{DOUBLE PRECISION} &  \texttt{esoft} &
value of $E_\mathrm{soft}$\\
\hline
\end{tabular}\\
The subroutine returns the value of $\Omega$, $Y$ and $E_\mathrm{soft}$.
Whenever a jet configuration is set up or modified,
the corresponding values of $\Omega$, $Y$ and $E_\mathrm{soft}$ 
are recalculated and can be retrieved using this subroutine.
\paragraph*{\texttt{get\_jet ( j, e, xta, phi, q, qtilde, ephys, qphys )}}~\\
\begin{tabular}{lll}
\hline
\emph{input:} & &\\
\texttt{INTEGER} &  \texttt{j} &
index of the jet\\
\hline
\emph{output:} & &\\
\texttt{DOUBLE PRECISION} &  \texttt{e} &
normalized energy \\
& & or normalized transverse energy\\
\texttt{DOUBLE PRECISION} &  \texttt{xta} &
angle $\theta_j$ or pseudorapidity $\eta_j$\\
\texttt{DOUBLE PRECISION} &  \texttt{phi} &
angle $\phi_j$\\
\texttt{DOUBLE PRECISION} &  \texttt{q(0:3)} &
normalized 4-momentum $q_j$\\
\texttt{DOUBLE PRECISION} &  \texttt{qtilde(0:3)} &
4-direction $\tilde{q}_j$\\
\texttt{DOUBLE PRECISION} &  \texttt{ephys} &
energy (or transverse energy)\\
& &  in physical units\\
\texttt{DOUBLE PRECISION} &  \texttt{qphys(0:3)} &
4-momentum $q_j$ in physical units\\

\hline
\end{tabular}\\
The subroutine returns parameters of the $j$-th jet, where $j$ obeys
$0\le j\le N$ and $j=0$ is the zeroth ``jet'',
name for the fractions of particles that do not belong to any jet
(i.e. soft energy).
For spherical kinematics the parameters are the usual energy, $E_j$,
normalized \texttt{e} and non-normalized \texttt{ephys}
(i.e. in the units of energy used in the input), the standard angles
$\theta_j$ (from the beam axis) and $\phi_j$.
For cylindrical kinematics the parameters are
the transverse energy, $E^\perp_j$,
normalized \texttt{e} and non-normalized \texttt{ephys}
(i.e. in the units of energy used in the input),
pseudorapidity $\eta_j$ and the standard angle $\phi_j$.
 All angles are in degrees.
For both kinematics the parameters \texttt{q(0:3)}
and \texttt{qtilde(0:3)} are the normalized and non-normalized
4-momentum of the jet. $\tilde{q}_j$ is the 4-direction defined in
section \ref{sec-def}.
\paragraph*{\texttt{get\_z ( a, z\_out )}}~\\
\begin{tabular}{lll}
\hline
\emph{input:} & &\\
\texttt{INTEGER} & \texttt{a} &
index of the particle\\
\hline
\emph{output:} & &\\
\texttt{DOUBLE PRECISION} & \texttt{z\_out(0:njets\_max)} &
components $z_{a0},z_{a1},...,z_{aN}$\\
\hline
\end{tabular}\\
The subroutine returns the components $z_{a0},z_{a1},...,z_{aN}$ of
the recombination matrix for the $a$-th particle. $a$ must satisfy
$1\le a\le n$.
The value of \texttt{z\_out(j)} is the $a$-th particle contribution
to the $j$-jet and $j=0$ corresponds to the soft energy.
Note: $\sum_\mathtt{j=0}^\mathtt{N}\mathtt{z\_out(j)}=1$.
\paragraph*{\texttt{get\_particle\_split ( a, total\_jets, jet, zj )}}~\\
\begin{tabular}{lll}
\hline
\emph{input:} & &\\
\texttt{INTEGER} &  \texttt{a} &
index of the particle\\
\hline
\emph{output:} & &\\
\texttt{INTEGER} &  \texttt{total\_jets} &
number of jets the particle\\
& & belongs to\\
\texttt{INTEGER} &  \texttt{jet(0:njets\_max)} &
indices of the jets\\
\texttt{DOUBLE PRECISION} &  \texttt{zj(0:njets\_max)} &
corresponding $z_{aj}$\\

\hline
\end{tabular}\\
For the $a$-th particle the subroutine returns:
the number of jets (including soft energy 0-th ``jet'')
which include a non-zero fraction of the particle ($z_{aj}\neq 0$),
the labels of the jets and the corresponding values of $z_{aj}$ in such
an order that
$\mathtt{zj\left(k\right)}\ge\mathtt{zj\left(j+1\right)}$. In other words,
the vector \texttt{zj(0:njets\_max)} is the collection of 
$z_{a0},z_{a1},...,z_{aN}$ ordered by their value (descending from
the left to right); only the components \texttt{(0:total\_jets-1)} are
different from zero.
\paragraph*{\texttt{get\_jet\_split ( j, nwhole, whole\_a, nfract, fract\_a, fract\_z )}}~\\
\begin{tabular}{lll}
\hline
\emph{input:} & &\\
\texttt{INTEGER} &  \texttt{j} &
index of the jet\\
\hline
\emph{output:} & &\\
\texttt{INTEGER} &  \texttt{nwhole} &
number of particles entirely\\
& & belonging to the jet\\
\texttt{INTEGER} &  \texttt{whole\_a(0:nparts\_max)} &
labels of the particles\\
\texttt{INTEGER} &  \texttt{nfract} &
number of particles\\
& & belonging in some fraction\\
\texttt{INTEGER} &  \texttt{fract\_a(0:nparts\_max)} &
labels of the particles\\
\texttt{DOUBLE PRECISION} &  \texttt{fract\_z(0:nparts\_max)} &
the fraction $z_{aj}$\\

\hline
\end{tabular}\\
For the $j$-th jet ($0\le j\le N$, $j=0$ is the soft energy)
the subroutine returns: 
\begin{itemize}
\item
number of the particles wholly in the jet, i.e. $z_{aj}=1$
\item
vector \texttt{whole\_a(0:nparts\_max)} with labels of such particles
(indices $a$)
\item
number of particles partially in the jet, i.e $0<z_{aj}<1$
\item
vector \texttt{fract\_a(0:nparts\_max)} with labels of such particles
(indices $a$)
\item
vector \texttt{fract\_z(0:nparts\_max)} with corresponding $z_{aj}$
for such particles.
\end{itemize}
The latter two vectors are synchronously ordered so that subsequent
components of \texttt{fract\_z(0:nparts\_max)} do not increase.
\subsubsection{Sample print routines}
\paragraph*{\texttt{print\_z\_raw}}~\\
is an example subroutine to print the recombination matrix
$\left\{z_{aj}\right\}$. A possible output may look like:
\begin{quote}
\begin{verbatim}
  a    background      1           2           3
 ---   ----------- ----------- ----------- -----------
   1    0.0000      0.0000      0.0000      1.0000
   2    0.0000      0.0000      0.0000      1.0000
   3    0.0000      0.0000      0.0000      1.0000
   4    0.0000      0.0000      0.0000      1.0000
 ...
\end{verbatim}
\end{quote}
\paragraph*{\texttt{print\_z\_nice}}~\\
is an example subroutine to print the recombination matrix
$\left\{z_{aj}\right\}$. A possible output may look like: 
\begin{quote}
\begin{verbatim}

 recombination matrix z by particle label a:

                          j e t   n u m b e r s
  a    background      1           2           3
 ---   ----------- ----------- ----------- -----------
   1        -           -           -          1.
   2        -           -           -          1.
   3        -           -           -          1.
   4        -           -           -          1.
 ...

\end{verbatim}
\end{quote}
%
%
%\item
\paragraph*{\texttt{print\_jets}}~\\
is an example subroutine to print properties of jets.
See the output of the example program in section \ref{sec-output}.
\paragraph*{\texttt{print\_particles}}~\\
is an example subroutine to print properties of particles.
A possible output may look like:
\scriptsize
\begin{quote}
\begin{verbatim}

 Configuration by particle:

                                        (soft energy is denoted as jet=0)
  a       E            E(%)    theta      phi     jet [ fraction ]; ...
 ---   -----------   -------   -------   -------   ---
   1    0.5100        6.7194   70.0000    0.0000     3
   2    0.4000        5.2701   90.0000    0.0000     3
   3    0.4000        5.2701   85.0000   10.0000     3
   4    0.2000        2.6350   84.0000  -10.0000     3
 ...
  24    0.2000        2.6350  170.0000   -7.0000     1
  25    7.0000E-02    0.9223   90.0000  170.0000     0
 ---------------------------
 TOTAL: 7.5900      100.0000
\end{verbatim}
\end{quote}
\normalsize                             
%
%
%\end{itemize}
%
%
%
%
%
%
\subsubsection{Example subroutine of straightforward jet search
 \texttt{Q\_search}}
\label{sec-q-search}
This is a simple jet search subroutine using \texttt{Q\_minimize}
as a key component.
It is possible that the user may want to modify it, 
for example, when trying to do something with local minima.
This subroutine uses only interface routines; 
it does not access internal data.

The subroutine tries to find 
the configuration of jets which minimizes $\Omega$ and ensures that 
$\Omega<\omega_\mathrm{cut}$ with the minimal number of jets 
(\texttt{njets}) starting from the number of jets 
previously set via \texttt{set\_njets\_start}
(usually the same for all events).
For each number of jets, the search is repeated \texttt{ntries} times,
each time with a different random initial value of
the recombination matrix $\left\{z_{aj}\right\}$
and the configuration with the lowest 
value of $\Omega$ is retained as a result.
Failure of the search is signaled by the condition \texttt{njets}=0.

Note that \texttt{Q\_search} randomizes the initial value of
$\left\{z_{aj}\right\}$, 
so it is meaningless to use it if one wants to specify 
the initial configuration for $\left\{z_{aj}\right\}$.
In this case, the user should use \texttt{Q\_minimize} directly. 
We comment that some other control options could be
to continue attempts until a specified number of attempts
fails to yield a better configuration or
to stop the search for new minimum if,
for example, the first three random initial
configurations yielded the same configurations
(the event has a single local minimum which is
automatically the global one; this is quite likely
and may be useful if CPU time is an issue).
\subsection{Compilation}
Optimal Jet Finder consists of the following files:
\begin{itemize}
\item \texttt{ojf\_014.f}
main file contains all subroutines and functions
\item \texttt{ojf\_com.fh}
contains definitions of common blocks
\item \texttt{ojf\_par.fh}
contains definitions of parameters
\item \texttt{ojf\_kin.fh}
contains definition of kinematics type parameters
\end{itemize}
The example programs
\texttt{example.f},
\texttt{ww160.f}, 
\texttt{ww160a.f}
with input or output files
\texttt{example.in},
\texttt{ww160.in},
\texttt{ww160.out},
\texttt{ww160a.out}
are added.

To compile and run any of example programs with OJF under Linux
equipped with g77 the user can type:
\begin{quote}
\begin{tabular}{ll}
\texttt{g77 user\_program.f ojf\_014.f -o executable\_file} & (enter)\\
\texttt{executable\_file} & (enter)
\end{tabular}                                  
\end{quote}
where \texttt{user\_program.f} is the name of the user own program applying
OJF. Each example program \texttt{example.f}, \texttt{ww160.f} or
\texttt{ww160a.f} can be used in its place. The files
\texttt{ojf\_com.fh}, \texttt{ojf\_par.fh}, and \texttt{ojf\_kin.fh}
should be available in the current directory (but not compiled).
\subsection{Example}
\label{sec-example}
The simplest possible example, file \texttt{example.f} below,
should give the idea how Optimal Jet Finder can be used.
The file \texttt{example.in} contains input data. Each line
corresponds to one particle and consists of
$E_a$, $\theta_a$ and $\phi_a$ for that particle.
The user is encouraged to study subroutine \texttt{Q\_search}
and programs \texttt{ww160.f}, \texttt{ww160a.f} providing
additional, more advanced examples.
\subsubsection{\texttt{example.f}}
\verb|      PROGRAM simplest_example|\\
\verb| |\\
\verb|      INCLUDE 'ojf_kin.fh'|\\
\verb| |\\
\verb|      DOUBLE PRECISION  Radius|\\
\verb|      DOUBLE PRECISION  e, theta, phi|\\
\verb|      DOUBLE PRECISION  o_fin, y_fin, e_fin|\\
\verb|      INTEGER           a, seed, nparts, njets, kinematics|\\
\verb|      LOGICAL           success|\\
\verb| |\\
\emph{number of jets is chosen}\\
\verb|      njets = 3|\\
\emph{seed for random generation of the recombination matrix}\\
\verb|      seed  = 13|\\
\emph{$R$ parameter from equation (\ref{def-omega}), section \ref{sec-algo}}\\ 
\verb|      Radius = 1.0|\\
\emph{choose spherical (lepton collisions) kinematics}\\
\verb|      kinematics = sphere|\\
\verb| |\\
\emph{file with input data is opened}\\
\verb|      OPEN(10, FILE='example.in', FORM='formatted', STATUS = 'old')|\\
\verb| |\\
\emph{input event setup starts}\\ 
\verb|      CALL event_setup_begin ( kinematics )|\\
\verb| |\\
\emph{loop over all particles in the event}\\
\verb|      nparts = 0|\\
\verb|      DO a = 1, 1999|\\
\verb|         READ(10,*, end=1000, err=1000) e, theta, phi|\\
\verb|         CALL add_particle ( e, theta, phi )|\\
\verb|         nparts = nparts + 1|\\
\verb|      ENDDO|\\
\verb|      |\\
\verb| 1000 CLOSE(10)|\\
\verb| |\\
\emph{input of the event ends}\\
\verb|      CALL event_setup_end|\\
\verb| |\\
\emph{set up random the initial value of the recombination matrix}\\
\verb|      CALL jets_setup_begin ( njets, Radius )|\\
\verb|      CALL set_seed ( seed )|\\
\verb|      CALL init_z_random_all|\\
\verb|      CALL jets_setup_end|\\
\verb| |\\
\emph{minimize $\Omega$}\\
\verb|      CALL Q_minimize ( success )|\\
\verb|      IF (.NOT. success) STOP 'minimum not found'|\\
\verb| |\\
\emph{get and print the values of $\;\Omega$, $Y$ and $E_\mathrm{soft}$
for the final jet configuration}\\
\verb|      CALL get_criterion ( o_fin, y_fin, e_fin )|\\
\verb| |\\
\verb|      WRITE(*,*) 'Omega  =', o_fin|\\
\verb|      WRITE(*,*) 'Y      =', y_fin|\\
\verb|      WRITE(*,*) 'E_soft =', e_fin|\\
\verb| |\\
\emph{prints properties of the resulting jets}\\
\verb|      call print_jets|\\
\verb| |\\
\verb|      END|\\
\subsubsection{Output of the example}
\label{sec-output}
\begin{verbatim}
 Omega  =  0.293404849
 Y      =  0.0338528071
 E_soft =  0.259552042  

 SPHERE:  3 jets processed

 Configuration by jet:

 jet       E           E(%)     theta      phi
 ---   -----------   -------   -------   -------
   1     1.380       18.1818  138.3848  -52.8876
   2     1.220       16.0738  124.4338  -26.6115
   3     3.020       39.7892   81.0226    0.3566
 ---------------------------
 TOTAL: 5.6200       74.0448

 Particle content by jet:

 jet label  1 (  3 particle(s) ):
         E(%)  =   18.18
        theta  =   138.4
          phi  =  -52.89
   3 particle(s) in jet as a whole:  21  22  24

 jet label  2 (  4 particle(s) ):
         E(%)  =   16.07
        theta  =   124.4
          phi  =  -26.61
   4 particle(s) in jet as a whole:  17  18  19  20

 jet label  3 (  8 particle(s) ):
         E(%)  =   39.79
        theta  =   81.02
          phi  =  0.3566
   8 particle(s) in jet as a whole:   1   2   3   4 
  5   6   7   8

 soft energy ( 10 particle(s) ):
  10 whole particle(s) in soft energy:   9  10  11 
 12  13  14  15  16  23  25
 no particles partially in soft energy
\end{verbatim}
%
%
%
%%%%%%%%%%%%%%%%%%%%%%%%%%%%%%%%%%%%%%%%%%%%%%
%
%
%
\section{Definitions of constants: \texttt{ojf\_par.fh}}
In this section we explain the meaning of the parameters defined
in the header file \texttt{ojf\_par.fh} and give their default
values.\\ \\
\texttt{INTEGER njets\_max}=50\\
The maximal number of jets;
used for example to define the size of matrices.\\ \\
\texttt{INTEGER nparts\_max}=2000\\
The maximal number of particles in the event;
used for example to define the size of matrices.\\ \\
\texttt{DOUBLE PRECISION zero}=0\\
\texttt{DOUBLE PRECISION one}=1\\
\texttt{DOUBLE PRECISION inf}=$10^{100}$\\
are the numerical constants.\\ \\
\texttt{DOUBLE PRECISION eps\_snap}=$10^{-3}$\\
If $z_{aj}<\mathtt{eps\_snap}$ than $z_{aj}$ is set to zero,
i.e. the particle is snapped to the boundary of the simplex.
The parameter is used in subroutines \texttt{z\_snap}
and \texttt{z\_assert}.\\ \\
\texttt{DOUBLE PRECISION eps\_round}=$10^{-6}$\\
\texttt{DOUBLE PRECISION eps\_sum}=$10^{-8}$\\
\texttt{DOUBLE PRECISION eps\_sum0}=$10^{-6}$\\
\texttt{DOUBLE PRECISION eps\_sum1}=$10^{-4}$\\
The constants are used to keep control of rounding errors.
If some variable exceeds the allowing range
of values more than \texttt{eps\_}, the error message
is generated and the program is terminated. The constants
are used in the subroutines \texttt{d\_minus\_snap},
\texttt{z\_snap}, \texttt{d\_assert}, \texttt{z\_assert},
\texttt{z\_force\_to\_simplex}
and in the function \texttt{pos\_prod}.\\ \\
\texttt{DOUBLE PRECISION eps\_norm}=$10^{-6}$\\
The constant is used to determine whether the norm
of the 3-vector $\mathbf{q}_j$ (or transverse part of the
norm in case of cylindrical dynamics) is zero.
It is used in the subroutine \texttt{j\_eval\_nonlinear}.\\ \\
\texttt{DOUBLE PRECISION eps\_Et}=$10^{-6}$\\
The constant is used to handle small values of
the transverse energy of a jet. It is used in
the subroutines \texttt{grad\_Y} and
\texttt{j\_eval\_nonlinear}.\\ \\
\texttt{DOUBLE PRECISION eps\_dist}=$10^{-6}$\\
See section \ref{sec-q-min}. The constant is used
to determine when to stop subsequent reductions of
the step $\tau$. The constant is used in
the subroutine \texttt{Q\_minimize\_wrt}.\\ \\
\texttt{DOUBLE PRECISION eps\_radius}=$10^{-3}$\\
The constant sets the limit on the smallest value
of $R$, the parameter from equation (\ref{def-omega}).
It is used in the subroutines \texttt{jet\_setup\_begin}
and \texttt{reset\_Radius}.\\ \\
\texttt{DOUBLE PRECISION inf\_step}=$10^{30}$\\
See section \ref{sec-q-min}. ``Infinite'' step means
that the particle should not be moved. The constant
is used in subroutines \texttt{Q\_minimize\_wrt},
\texttt{d\_eval\_step} and\\ \texttt{z\_move\_by}.\\ \\
It is imaginable that some of the parameters above
may need to be changed but the user is advised to be careful
when doing this. In particular, smaller values of some parameters
would enhance sensitivity to rounding errors,
causing the safety checks to generate error messages and terminate
the program.
One may change \texttt{eps\_snap} to a smaller value, say $10^{-5}$,
and see if the results would change; for a small fraction of events this may
slow the finding of jets but help to better identify local minima.\\ \\
\texttt{INTEGER random\_m}=259200\\
The constant is used by the random number generator, the subroutine
\texttt{seed} and the function \texttt{random()}.
\\ \\
The constants below play a technical role and are not supposed to be
changed. The reason for defining them is cleared in the next section.\\ \\
\texttt{INTEGER par\_Et}=4\\
\texttt{INTEGER par\_eta}=5\\
\texttt{INTEGER par\_Eteta}=6\\
\texttt{INTEGER par\_y}=7\\
\texttt{INTEGER par\_p0shmpzch}=8\\
\texttt{INTEGER par\_tilde}=9\\
\section{Common block definitions: \texttt{ojf\_com.fh}}
The header file \texttt{ojf\_com.fh} contains common block definitions
and is included in most of the subroutines. The user is not
supposed to write to common blocks directly but to use interface
subroutines. Data that cannot be accessed that way is not supposed
to be used by the user.\\
\subsection{Input of the event}
\begin{verbatim}
        COMMON /ojf_event/
     &      ojf_event_begin, ojf_event_set, 
     &      ojf_kinematics, ojf_nparts,
     &      ojf_p, ojf_e, ojf_e_scale
\end{verbatim}
\texttt{LOGICAL ojf\_event\_begin, ojf\_event\_set}\\
The two logical values bracket the event setup:\\
\texttt{FALSE, FALSE} - at start of program, no event has been set up;\\
\texttt{TRUE , FALSE} - event setup in progress, adding particles;\\
\texttt{FALSE, TRUE } - event setup completed, can search for jets.\\ \\
\texttt{INTEGER ojf\_kinematics}\\
The variable marks the type of kinematics: 1 - spherical kinematics
of lepton collisions, 2 - cylindrical kinematics of hadron collisions.\\ \\
\texttt{INTEGER ojf\_nparts}\\
The number of particles in the event.\\ \\
\texttt{DOUBLE PRECISION ojf\_p(0:6, 1:nparts\_max)}\\
The matrix stores the properties of particles:\\
\begin{tabular}{ll}
\hline
\texttt{ojf\_p(0, particle\_label)} & energy $E_a$\\
\texttt{ojf\_p(1, particle\_label)} & x-component of momentum $p_a$\\
\texttt{ojf\_p(2, particle\_label)} & y-component of momentum $p_a$\\
\texttt{ojf\_p(3, particle\_label)} & z-component of momentum $p_a$\\
\texttt{ojf\_p(4, particle\_label)} & transverse energy $E^\perp_a$\\
\texttt{ojf\_p(5, particle\_label)} & pseudorapidity $\eta_a$\\
\texttt{ojf\_p(6, particle\_label)} & combination $E^\perp_a\cdot\eta_a$\\
\hline
\end{tabular}\\
and \texttt{particle\_label} is the index $a$ of the particle. The constants
\texttt{par\_Et}=4, \texttt{par\_eta}=5, \texttt{par\_Eteta}=6 are
defined to access the components of the matrix,\\
e.g. \texttt{ojf\_p(par\_Eteta, particle\_label)}.\\ \\
\texttt{DOUBLE PRECISION ojf\_e(1:nparts\_max)}\\
The vector stores the energies of the particles.\\ \\
\texttt{DOUBLE PRECISION ojf\_e\_scale}\\
The variable stores the energy scaling factor (see section
\ref{normalization}).
\subsection{Configuration of jets (output)}
\begin{verbatim}
        COMMON /ojf_jets/
     &      ojf_jets_begin, ojf_jets_set,
     &      ojf_njets, ojf_seed, ojf_Radius,
     &      ojf_z, ojf_b, ojf_q, 
     &      ojf_Omega, ojf_Y, ojf_Esoft
\end{verbatim}
\texttt{LOGICAL ojf\_jets\_begin, ojf\_jets\_set}\\
The two logical values bracket setup of initial jet configuration:\\
\texttt{FALSE, FALSE} - at start of program, or after event set up;\\
\texttt{TRUE,~~FALSE} - jets setup in progress, change anything;\\
\texttt{FALSE, TRUE } - jets setup complete, can do minimization.\\ \\
\texttt{INTEGER ojf\_njets}\\
The number of jets in the current configuration.\\ \\
\texttt{INTEGER ojf\_seed}\\
The seed used to generate the current (random) jet configuration.\\ \\
\texttt{DOUBLE PRECISION ojf\_Radius}\\
The value of $R$, the parameter in equation (\ref{def-omega}).\\ \\
\texttt{DOUBLE PRECISION ojf\_z(0:njets\_max,1:nparts\_max)}\\
The recombination matrix, $\left\{z_{aj}\right\}$.\\ \\
\texttt{LOGICAL ojf\_b(0:njets\_max,1:nparts\_max)}\\
It is used to indicate that the particle belongs to (\texttt{TRUE})  
or does not belong (\texttt{FALSE}) to the boundaries of the simplex,
i.e. $z_{aj}=0$.\\ \\
\texttt{DOUBLE PRECISION ojf\_q(0:12, 1:jets\_max)}\\
The matrix stores the properties of particles:\\
\begin{tabular}{ll}
\hline
\texttt{ojf\_q( 0, jet\_label)} & energy $E_j$\\
\texttt{ojf\_q( 1, jet\_label)} & x-component of momentum $q_j$\\
\texttt{ojf\_q( 2, jet\_label)} & y-component of momentum $q_j$\\
\texttt{ojf\_q( 3, jet\_label)} & z-component of momentum $q_j$\\
\texttt{ojf\_q( 4, jet\_label)} & transverse energy $E^\perp_j$\\
\texttt{ojf\_q( 5, jet\_label)} & pseudorapidity $\eta_j$\\
\texttt{ojf\_q( 6, jet\_label)} & combination $E^\perp_j\cdot\eta_j$\\
\texttt{ojf\_q( 7, jet\_label)} & fuzziness $Y$\\
\texttt{ojf\_q( 8, jet\_label)} 
& combination $\left(q_j\right)^0\cdot\left(\tilde{q}_j\right)^3
-\left(q_j\right)^3\cdot\left(\tilde{q}_j\right)^0$\\
\texttt{ojf\_q( 9, jet\_label)} & 0-component of 4-direction $\tilde{q}_j$\\
\texttt{ojf\_q(10, jet\_label)} & x-component of 4-direction $\tilde{q}_j$\\
\texttt{ojf\_q(11, jet\_label)} & y-component of 4-direction $\tilde{q}_j$\\
\texttt{ojf\_q(12, jet\_label)} & z-component of 4-direction $\tilde{q}_j$\\
\hline
\end{tabular}\\
\texttt{jet\_label} is the index $j$ of the jet. The constants
\texttt{par\_Et}=4, \texttt{par\_eta}=5, \texttt{par\_Eteta}=6,
\texttt{par\_y}=7, \texttt{par\_p0shmpzch}=8, \texttt{par\_tilde}=9
are defined to access the components of the matrix, e.g.
\texttt{ojf\_q(par\_y, jet\_label)}.\\ \\
\texttt{DOUBLE PRECISION ojf\_Omega, ojf\_Y, ojf\_Esoft}\\
The variables store the values of $\Omega$, $Y$ and $E_\mathrm{soft}$,
see equation (\ref{def-omega}).\\ \\
The remaining common block \texttt{/ojf\_work/} contains the definitions
of ``work'' variables, mainly of the types explained above.
The ``work'' variables are used at the intermediate stages of computations.
\section{Conclusion}
We implemented an algorithm for minimization of $\Omega$,
the criterion for finding the optimal jet configuration.
The user is provided with the example program and
subroutine \texttt{Q\_search} as the simplest possible
scenarios of jet finding. It is not inconceivable that
practical applications will require more sofisticated
ways of using the provided library. To facilitate this,
we supplied the user with numerous interface subroutines
allowing for easy control of the program.

We did not explore all possible ideas for optimizing
of jet search as we wanted to leave the program generic.
For example, in some cases it is possible to apply OJF
in a few stages. First, apply it to particles with the highest
energies, then use the result as the initial value of
the recombination matrix for the next stage at which
the rest of particles are added (or some next portion).
It may increase the speed of finding jets without
significant deterioration of the results [1].
However this depends on the particular physical process at hand.
%
%
%%%%%%%%%%%%%%%%%%%%%%%%%%%%%%%%%%%%%%%%%%%%%%
%
%
%

\end{document}